\newcommand{\be}{\begin{equation}}
\newcommand{\ee}{\end{equation}}
\documentstyle[12pt]{article}
\newlength{\dinwidth}
\newlength{\dinmargin}
\begin{document}

\title{
\hfill {\large DTP/94/30}\\
\hfill {\large May 1994}\\
\vspace {1cm}
{ \bf Transverse energy flow at HERA}
\\[30pt]
\author{\bf 
K.~Golec\--Biernat$^\ast$, 
J.~Kwieci\'nski$^\ast$, A.D.~Martin$^\dagger$ and
P.J.~Sutton$^\ddagger$
\\[20pt]
{$^\ast$ Henryk Niewodnicza\'nski Institute of Nuclear Physics,}
\\[3pt]
{ul. Radzikowskiego 152, 31-342 Krak\'ow, Poland } 
\\[20pt]
{$^\dagger${Dept. of Physics, University of Durham, DH1 3LE, England}}
\\[20pt]
{$^\ddagger${Dept. of Physics, University of Manchester, M13 9PL, England}}
\\[30pt]
       }
\date{}
     }

\maketitle

\begin{abstract}
 {\large  We calculate the transverse energy flow accompanying
small $x$ deep-inelastic events and compare with recent data 
obtained at HERA.  In the central region between the current jet
and the remnants of the proton we find that BFKL leading
$\ln(1/x)$ dynamics gives a distinctively large transverse
energy distribution, in approximate agreement with recent data.}
\end{abstract}

\thispagestyle{empty}

\newpage
\setcounter{page}{1}
\vspace{1 cm}

     The structure function $F_2(x,Q^2)$ for deep-inelastic
electron-proton scattering has recently been measured \cite{HERA} in the
small $x$ region accessible at HERA, $x\sim 10^{-3}$.
The measured values show a striking rise with decreasing
$x$ which is entirely consistent with perturbative QCD expectations
based on the precocious onset of the Balitsky-Fadin-Kuraev-Lipatov 
(BFKL) \cite{BFKL} leading 
$\ln (1/x)$ behaviour. However, the observed small
$x$ behaviour of $F_2$ can equally well be mimicked by conventional
dynamics based on Altarelli-Parisi (GLAP) evolution \cite{AGKMS}, 
where the steep 
behaviour is either put into the starting distributions or,
alternatively, generated by $Q^2$ evolution from a very low
scale $Q_0^2$.  

   To obtain a sensitive discriminator between
BFKL and conventional GLAP dynamics we need to look into the
properties of the final state.  A relevant observable, which has
been recently measured at HERA by the H1 collaboration \cite{H1}, is the
transverse energy ($E_T$) flow in deep-inelastic events. 
The deep-inelastic data hint at an excess of $E_T$ in the forward part of
the central region when compared to Monte Carlo simulations
which incorporate GLAP evolution.   GLAP evolution corresponds to
a summation of large $\ln Q^2$ terms, which is equivalent (in a
physical gauge) to the summation of ladder diagrams with strongly
ordered transverse momenta ($k_T$) of the emitted partons along the
ladder: that is  $Q^2\gg k^2_n\gg ...\gg k^2_1$, where we have omitted
the subscript $T$ on $k^2_{Ti}$.   

At small $x$ it becomes necessary to resum the large $\ln (1/x)$ 
terms and this is accomplished via the BFKL equation.
The gluon `ladder' diagrams relevant to this equation do not
have the strong-ordering of the transverse momenta
that is present in the GLAP diagrams.
As a result more transverse energy is expected in the central
region (between the current jet and the proton remnants) than
would occur from conventional GLAP dynamics.  These expectations are
confirmed by explicit calculations \cite{KMSG}. They are also hinted 
at by Monte Carlo simulations (which incorporate BFKL effects)
of the gluon radiation accompanying heavy quark
production at sufficiently small $x$, see Fig.~8 of ref.~\cite{MW} 
or Fig.~9(b) of ref.~\cite{WEBBER}. 

In ref.~\cite{KMSG} we were mainly concerned with formalism and in 
using analytic methods to gain
insight into the general characteristic features of the BFKL
description of the energy flow in the small
$x$ regime.  For fixed $\alpha_s$ we derived an analytic form
of the $E_T$ flow in the central region of deep-inelastic
events at small $x$. 
The $E_T$ distribution was found to be a broad Gaussian-shaped plateau
as a function of rapidity, with a height that increases
with decreasing $x$, and/or increasing $Q^2$.  We also performed 
numerical estimates of the $E_T$ flow, which included the
effects of running $\alpha_s$.  These, more physical, calculations
qualitatively confirmed the characteristic features of the fixed-$\alpha
_s$ treatment, but did not fully cover 
the central region.  The BFKL gluon emissions are only one
source of transverse energy.  There will also be contributions
arising from
parton radiation from the current jet and the proton remnants,
and some enhancement of the $E_T$ flow from the subsequent
hadronization.  Here, in order to
attempt a realistic comparison with the recent data, we extend the
BFKL-based 
calculations of $E_T$ to cover a larger interval of rapidity and
we use a Monte Carlo to simulate the effects
of radiation from the current jet and of hadronization.
We choose the 
LEPTO Monte Carlo \cite{LEPTO}, as it gives
a good description of final state observables in 
deep-inelastic scattering (and $e^+ e^-$ collisions) in
regions insensitive to BFKL small $x$ dynamics.
The matrix element (ME) + parton showering (PS)
structure of the programme has the advantage that the 
GLAP-based initial state radiation can be isolated and therefore,
in principle, be substituted by BFKL gluon emissions.

    The energy flow accompanying deep-inelastic events, in a
small interval about $x,Q^2$, is given by \cite{KMSG}
\be
 {{\partial E_T}\over {\partial \ln (1/x_j)}}\simeq {1\over F_2} \int
 dk_j^2 \left | {\bf k_j} \right | \int {d^2k_p\over \pi k^4_p}
 \int {d^2k_{\gamma}\over k^4_{\gamma}} \left ( {3\alpha_s \over
\pi} {k^2_p k^2_\gamma\over k^2_j} \right ) {\cal F}_2(x/x_j,
k^2_\gamma,Q^2)f(x_j,k_p^2) \delta^{(2)}(k_j-k_{\gamma}-
k_p)
\label{eq:a}
\ee
where the transverse momenta are defined in Fig.1(a).  For simplicity
we have omitted the longitudinal structure function and
assumed that $F_2=2xF_1$.  It is straightforward to include
the small correction arising from $F_L=F_2-2xF_1$. 
The function $f$ is the unintegrated
gluon distribution of the proton in which the $k^2_p$ integration
 is unfolded.  To be precise
\be
 x_{j}g(x_j,\mu^2) = \int^{\mu^2} {dk^2_p\over k^2_p} f(x_j,k^2_p,\mu^2)
\label{eq:b}
\ee
gives the conventional gluon density at a scale $\mu^2$.  
In the leading $\ln(1/x_j)$ approximation $f$ is found to be independent of
the scale $\mu^2$ \cite{ITAL} and for this reason we have omitted the scale
variable from the arguments of $f$ in (\ref{eq:a}). 
In this small $x_j$ approximation 
the function $f$ satisfies a BFKL equation which
effectively sums the soft gluon emissions below the emitted
$(x_j,k_j)$ gluon in Fig.1(a).  
The same remarks apply to ${\cal F}_2$ such that for
sufficiently small $x/x_j$, the function ${\cal F}_2$ 
becomes scale independent and satisfies
a BFKL equation which effectively sums the soft gluon
emissions above the emitted gluon.

    As in ref.~\cite{KMSG}, $f(x_j,k_p^2)$ is determined for $x_{j}
<10^{-2}$ by step-by-step integration of the BFKL equation
down in $x_j$ starting from a gluon distribution at $x_{j}=10^{-2}$
obtained from the MRS set of partons of ref.~\cite{MRS},
and ${\cal F}_2$ is calculated for $x/x_j<10^{-1}$, as in 
ref.~\cite{KMS}, by
evolving down from the quark box (and crossed box)
contribution, ${\cal F}_2^{(0)}$, evaluated at $x/x_{j}=10^{-1}$.
Even in the lowest $x$ region accessible at HERA, $x\sim 10^{-4}$,
the above QCD prediction is limited to the $x_j$ interval
$10^{-3}<x_{j}<10^{-2}$.  
In Fig.~2(a) the continuous curve for $10^{-3} < x_j < 10^{-2}$ shows the
$E_T$ distribution for $x=10^{-4}$ and $Q^2 = 10$ GeV$^2$ 
-- values which are representative 
of the lowest $x$ regime accessible for deep-inelastic scattering
at HERA. The infrared cut-off on the transverse momentum integrations
is taken to be $k_0^2 =1$ GeV$^2$ throughout, a value for which 
the calculated \cite{AKMS} values of $F_2$ are consistent with the recent
measurements at HERA \cite{HERA}.  The predictions for the $E_T$
flow are less sensitive to the choice of the cut-off than those
for $F_2$.  The ultraviolet cut-off is chosen to be $Q^2 /z$ as
implied by energy-momentum conservation \cite{FHS}, where
$z=x/x_j$. 

To extend the prediction of the $E_T$ flow into the region $x_j > 10^{-2}$ we
proceed as follows. First we continue to use formula (\ref{eq:a}), but with 
\be
\left .
f(x_j,k_p^2) = {\partial (x_j g(x_j,\mu^2)) \over \partial \ln \mu^2} 
\right |_{\mu^2 = k_p^2}
\label{eq:c}
\ee
with the gluon taken from ref.~\cite{MRS}. This expression for $f$ 
follows from (\ref{eq:b}), since, in the leading $\ln(1/x_j)$ limit, $f$ is
independent of $\mu^2$. However, as $x_j$ increases we soon
reach the stage when the scale dependence of $f(x_j,k_p^2,\mu^2)$ can no
longer be neglected 
and so (\ref{eq:c}) becomes invalid. The gluon distribution 
$f$ should always be positive, whereas the logarithmic derivative of $xg$
becomes negative with increasing $x_j$ due to the increasing importance of
the usual virtual corrections of the GLAP equation. 
For this reason (as well as the omission of a quark 
contribution) the $E_T$ flow calculated from the above prescription
will be a larger and larger underestimate as $x_j$ increases
above $x_j \simeq 10^{-2}$. This effect
can be clearly seen in Fig.~2.

A better approach for these larger values of $x_j$ is to use 
the usual strong ordering at the
gluon emission vertex, $k^2_j\gg k^2_p$, so that
$k^2_{\gamma}\approx k^2_j$.  Then, on making use of (\ref{eq:b}), 
eq.(\ref{eq:a}) simplifies to
\be
{\partial E_T\over {\partial \ln (1/x_j)}}={1\over F_2}\int
 {dk^2_j\over k^4_j}{3\alpha_s\over \pi}\left |
{\bf k_j} \right | x_j\sum_a f_{a}(x_j,k_j^2)
{\cal F}_{2}(x/x_j,k_j^2,Q^2),
\label{eq:d}
\ee
where the ``effective'' parton combination $\sum_a f_a\equiv
g+\frac{4}{9}(q+\bar q)$ arises from the dominance of
gluon exchange \cite{CJM}.  In this way we include
the contributions when parton $a$ of Fig.1(b) is either a
quark or an antiquark, as well as the gluonic component which
was dominant in the smaller $x_j$ region.
For sufficiently large $x_j$ (but away from the proton remnants) formula 
(\ref{eq:d})
will give a much more reliable prediction for the transverse energy flow,
but as $x_j$ decreases the strong-ordering assumption becomes less valid and
the consequent neglect of regions of phase space means that the $E_T$ flow
will again be underestimated. This is apparent from Fig.~2(a) which shows the
predictions of (\ref{eq:d}) as a continuous curve in the 
interval $0.01 < x_j < 1$. 
In Fig.~2(b) we show the results for $x=5.7 \times 10^{-4}$
and $Q^2 = 15$ GeV$^2$. This choice represents the average values \cite{ADRG}
of the variables for  
the $E_T$ distribution, 
accompanying the deep-inelastic events with $x<10^{-3}$,
which was observed by the H1 collaboration \cite{H1}. 
Again there is a reasonably flat plateau in the central region,
but about $0.4$ GeV lower than that of Fig.~2(a) and covering a smaller
rapidity interval on account of the larger value of $x$. 
In summary, in the region 
$10^{-2} \stackrel{<}{\sim} x_j \stackrel{<}{\sim} 10^{-1}$ the $E_T$ flow is
underestimated by both (\ref{eq:a}) and (\ref{eq:d}), 
but at different ends of the interval.
Formula (\ref{eq:a}) is valid for $x_j \stackrel{<}{\sim} 10^{-2}$, 
while formula (\ref{eq:d}) is reliable
for $x_j \stackrel{>}{\sim} 10^{-1}$. From the combination 
of the results shown in Fig.~2 we conclude that 
BFKL radiation (which accompanies deep-inelastic events) 
gives rise to an approximately flat $E_T$ distribution
in the central region with 
a height, which increases slowly with decreasing $x$, of about
2 GeV per unit of rapidity, in the HERA small $x$ regime.

We also performed the above calculations using GLAP evolution along the
ladders. Figure 2 shows that, as expected, this evolution
gives a much smaller transverse energy flow than BFKL evolution.
The discontinuity
in the GLAP results at $x_j = 10^{-2}$ is due to the inclusion of the 
quark distributions in the calculation for the 
large $x_j$ region. As expected the
quarks have little importance at small $x_j$ where the gluon dominates.

It is useful to translate the energy flow $\partial E_T/
\partial \ln x_j$ into a distribution in terms of rapidity, $y$ in the
HERA frame. First we note that 
in the virtual photon-proton centre-of-mass (cm) 
frame the rapidity is given by
\be
 y({\rm cm}) =\frac{1}{2} \ln \left ( x_j^2 Q^2\over {xk_j^2} \right ).
\label{eq:e}
\ee
In regions where the distribution is reasonably flat, a good estimate
of the $y$ distribution is obtained if we insert the local average
value of $k_j^2$ into (\ref{eq:e}).  
Secondly we translate from the virtual photon-proton cm frame to the
HERA laboratory frame using the formula

\be
 y  - y({\rm cm}) \simeq {1\over 2} \ln \left ({4xE_p^2 \over Q^2}
\right )
\label{eq:f}
\ee
which is valid at small $x$, since then the frames are approximately collinear.
$E_p$ is the energy of the incoming proton in the HERA
frame.  

At this stage our numerical results, shown in Fig.~2(b),
cannot be compared directly with the H1 data \cite{H1}. 
The distributions of Fig.2
correspond to $E_T$ arising from gluons radiated from the
initial state.  Admittedly, at small $x$, 
this is expected to be the dominant
effect in the central region between the current jet and the
proton remnants --  the large energy flow that is predicted 
is  characteristic of BFKL dynamics.  However the
calculation does not take into account the current jet and its
associated radiation, nor does it include any effects of hadronization.
We therefore need to estimate the importance of the various 
components contributing to the transverse energy flow. 
To this end we compare 
in Fig.3 the H1 measurement of the $E_T$ flow \cite{H1} with
four different distributions obtained using the LEPTO Monte Carlo
\cite{LEPTO}. The parameters of LEPTO have been 
tuned to the hadronization resulting from jet
production in $e^+ e^-$ collisions and, to some extent,
tuned to EMC deep-inelastic data in the higher $x$
regime \cite{ING}.
The four $E_T$ distributions resulting from LEPTO 
show the energy flow obtained (i) from
only parton showers radiated from the current jet, (ii) 
with parton showers from
the initial state incorporated,
(iii) from only parton showers from the current jet but with hadronization
effects included, and (iv) from parton showers from both the current jet and
the initial state together with hadronization. 
We see that the Monte Carlo gives 
a good description of the current jet and its associated
radiation, but that effects of initial state radiation and of
hadronization are unable to give sufficient $E_T$ in the
forward part of the central region at small $x$. 
This has prompted a further study of
the LEPTO Monte Carlo in
order to assess whether this deficiency at small $x$ is genuine
or if it can be tuned away \cite{ING}. This requires
looking in detail at the LEPTO modelling of the GLAP-based initial state
parton showers, and of hadronization of proton remnants more
complicated than those arising from 
a diquark. We note from ref.~\cite{H1} that a Monte Carlo based on the colour
dipole model appears to give a better description of the $E_T$
flow in the central region than does LEPTO.  On the other hand the
colour dipole model is less successful than LEPTO in describing 
energy-energy correlations \cite{H1} and the $Q^2$ dependence of 
the jet rates \cite{H1J}.  It remains to be seen
how much is simply parameter tuning and how much is directly
attributable to QCD dynamics in the various Monte Carlo
simulations.   

In Fig.~4 we confront the BFKL predictions with the HERA measurements
of the $E_T$ flow in the central region with 
$x < 10^{-3}$. The average values of the deep-inelastic
variables for these data correspond to 
$x=5.7 \times 10^{-4}$ and $Q^2 = 15$ GeV$^2$
\cite{ADRG}. 
The BFKL predictions are thus 
simply those of Fig.~2(b), but shown now in terms of
$y$, the rapidity in the HERA frame. This translation is achieved
via eqs~(\ref{eq:e}) and
(\ref{eq:f}). 
However, the comparison of the observed $E_T$ flow at small
$x$ with the BFKL-based estimates is clearly incomplete. As emphasized
above, we have omitted the effects of hadronization and of radiation from 
the current jet. The magnitude of these effects can be estimated from 
the LEPTO Monte Carlo. The appropriate histogram of Fig.~3 is
reproduced in Fig.~4. 

To obtain a first estimate of
the total $E_T$ flow we could simply add the $E_T$ resulting from
the BFKL emissions to the LEPTO distribution.  In other words the
hadronization effects, which in LEPTO arise from the colour
string stretching from the current quark to the diquark
proton remnants, are assumed to give an underlying
rapidity plateau whose gross features are insensitive
to the properties of the initial state radiation.
The Monte Carlo results in Fig.~3 give support for this assumption, since
we see that the level of hadronization is approximately independent of
whether or not we include parton showers from the initial state.

If such a straightforward addition were to be performed on the 
results shown in Fig.~4 then the total estimate of the $E_T$ flow 
would be in approximate agreement with the data. 
There are two caveats.  First, there may be some danger of
double counting of radiation from the current jet and,
second, our simple additive treatment of hadronization may be too na{\" \i}ve.
However, the first problem does not effect the central region,
and secondly hadronization effects appear to be much less than the
BFKL signal.

    To conclude, we have shown that small $x$ deep-inelastic scattering
is accompanied by a large $E_T$ flow in the central region
arising from soft gluon radiation.  This is a hallmark of
BFKL dynamics and arises from the relaxation of the strong-ordering of
transverse momenta.  The first experimental measurements of the
$E_T$ flow in small $x$ deep-inelastic events indicate that
there is significantly more $E_T$ than is
given by conventional QCD cascade models based on Altarelli-Parisi
evolution. Instead we find that they are in much better agreement 
with estimates which
incorporate BFKL evolution.  The latter dynamics are characterised by an
increase of the $E_T$ flow in the central region with decreasing
$x$.  Measurements of the energy flow in
different intervals of $x$, in the small $x$ regime, should therefore
allow a definitive check of the applicability of BFKL dynamics.

\vspace{1cm}

\noindent {\large\bf Acknowledgements}

  We thank Gunnar Ingelman, Albert de Roeck 
and Jacek Turnau for valuable discussions.
 Two of us (KGB, PJS) thank the Polish
KBN - British Council collaborative research programme for partial support.
This work has also been supported in part by Polish KBN grant 
no. 2 0198 91 01, by the UK Science and
Engineering Research Council and the EU under contract
no. CHRX-CT92-0004.


\newpage

\noindent{\Large \bf Figure Captions}
\begin{itemize}
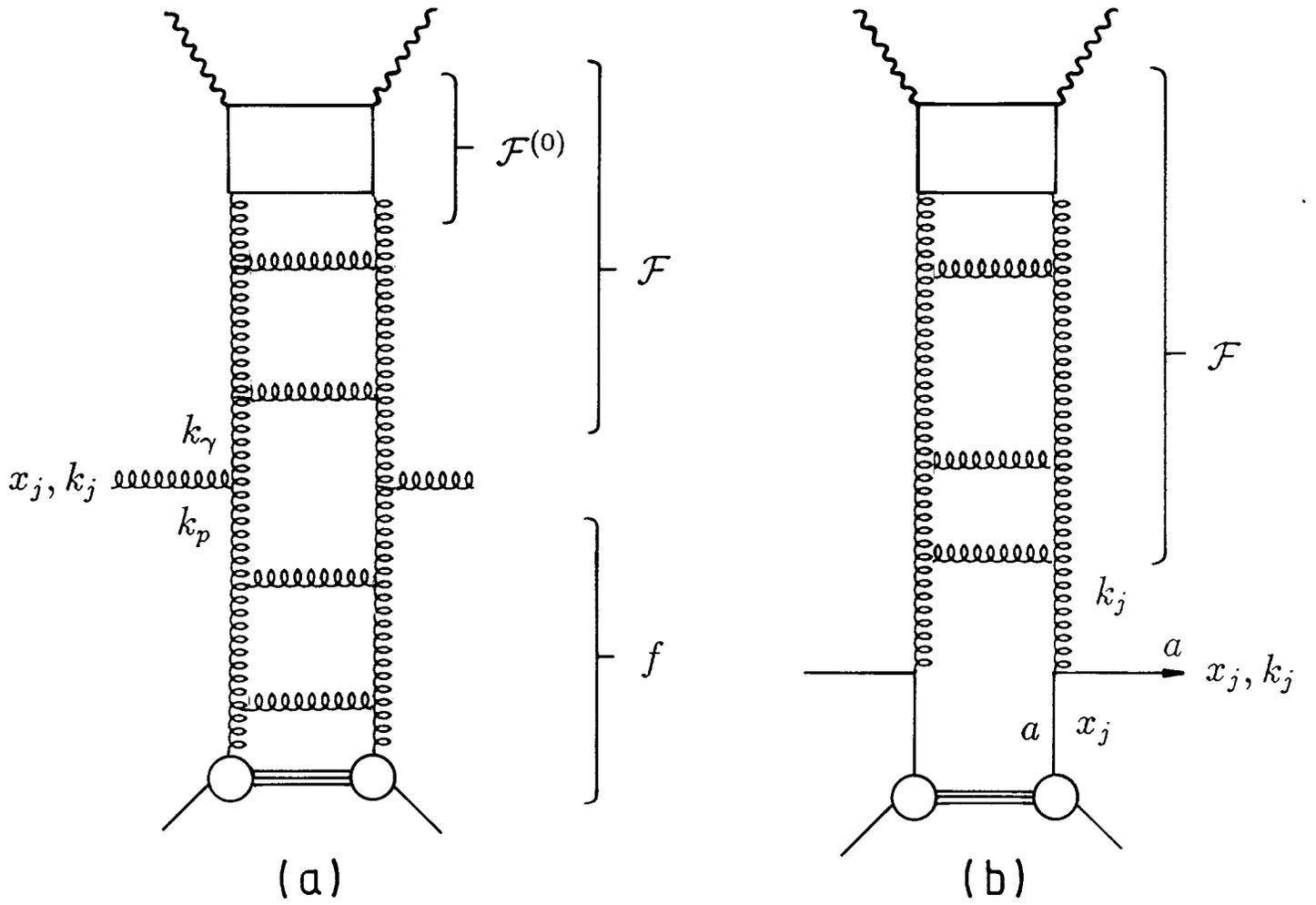
\item[Fig.\ 1:]
 (a) Diagrammatic representation of formula (\ref{eq:a}) showing the gluon
ladders which are resummed by the BFKL equations for $f$ and
${\cal F}_2$. (b) The representation of formula (\ref{eq:d}), which is obtained
by the simplification of (\ref{eq:a}) when $x_j$ is large; there is now 
strong-ordering at the (parton $a$)--gluon vertex.

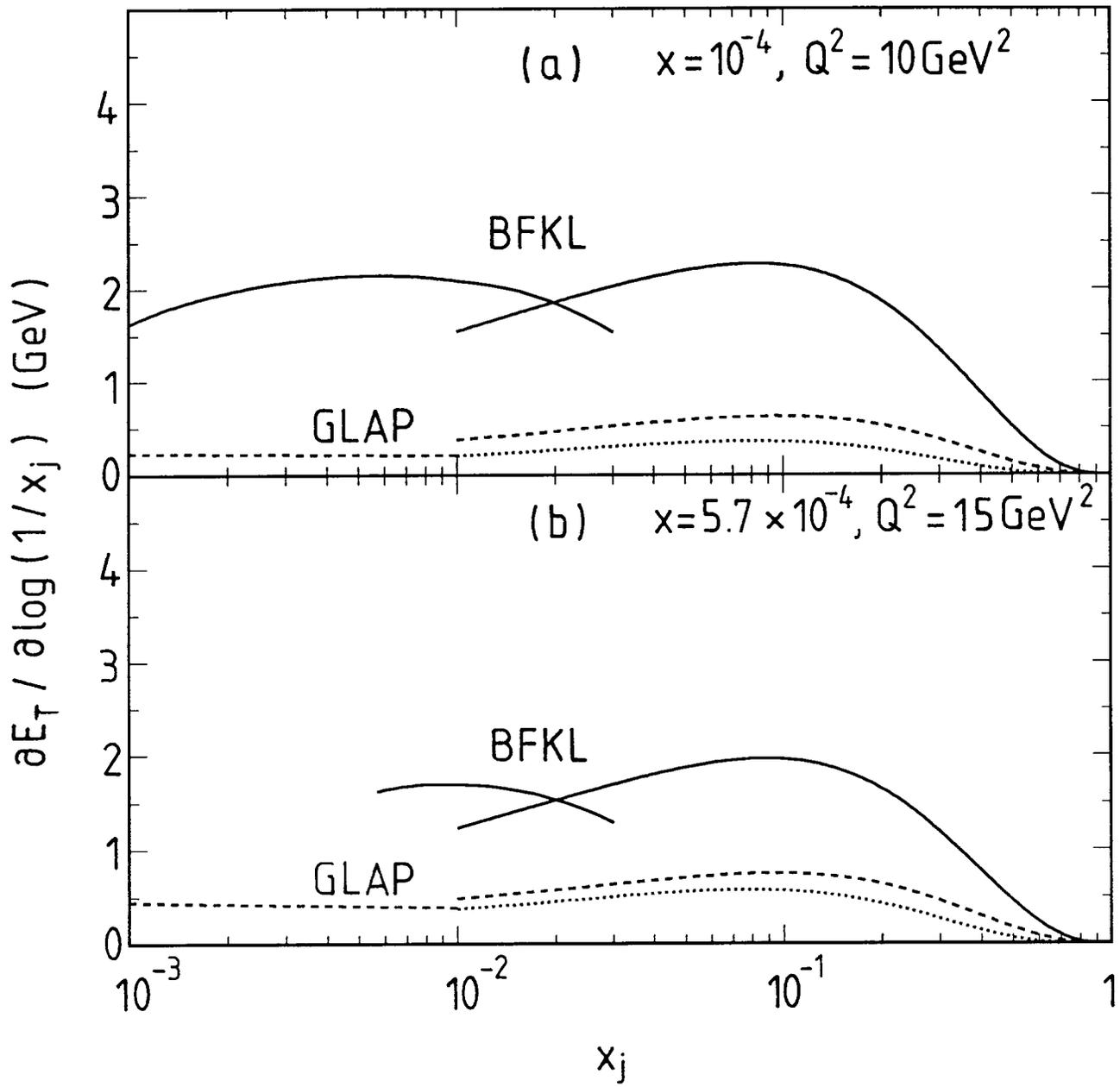
\item[Fig.\ 2:]
 Numerical calculation of the $E_T$ flow as a function of $x_j$ for
(a) $x=10^{-4}, Q^2=10$ GeV$^2$ and (b) $x=5.7 \times 10^{-4}, Q^2=15$
GeV$^2$. The latter choice of variables is
relevant to the HERA data for $x<10^{-3}$.
The effects of the current jet (and its associated radiation)
and of hadronization are not included. The continuous curves are 
based on BFKL dynamics: formula (\ref{eq:a}) is 
used for $x_j<10^{-2}$ and formula
(\ref{eq:d}) is used for $x_j > 10^{-2}$. 
For comparison, the dashed curves
show the $E_T$ flow calculated using GLAP evolution. The
dotted curve shows the effect of including only gluons at large $x_j$
in the GLAP evolution.  

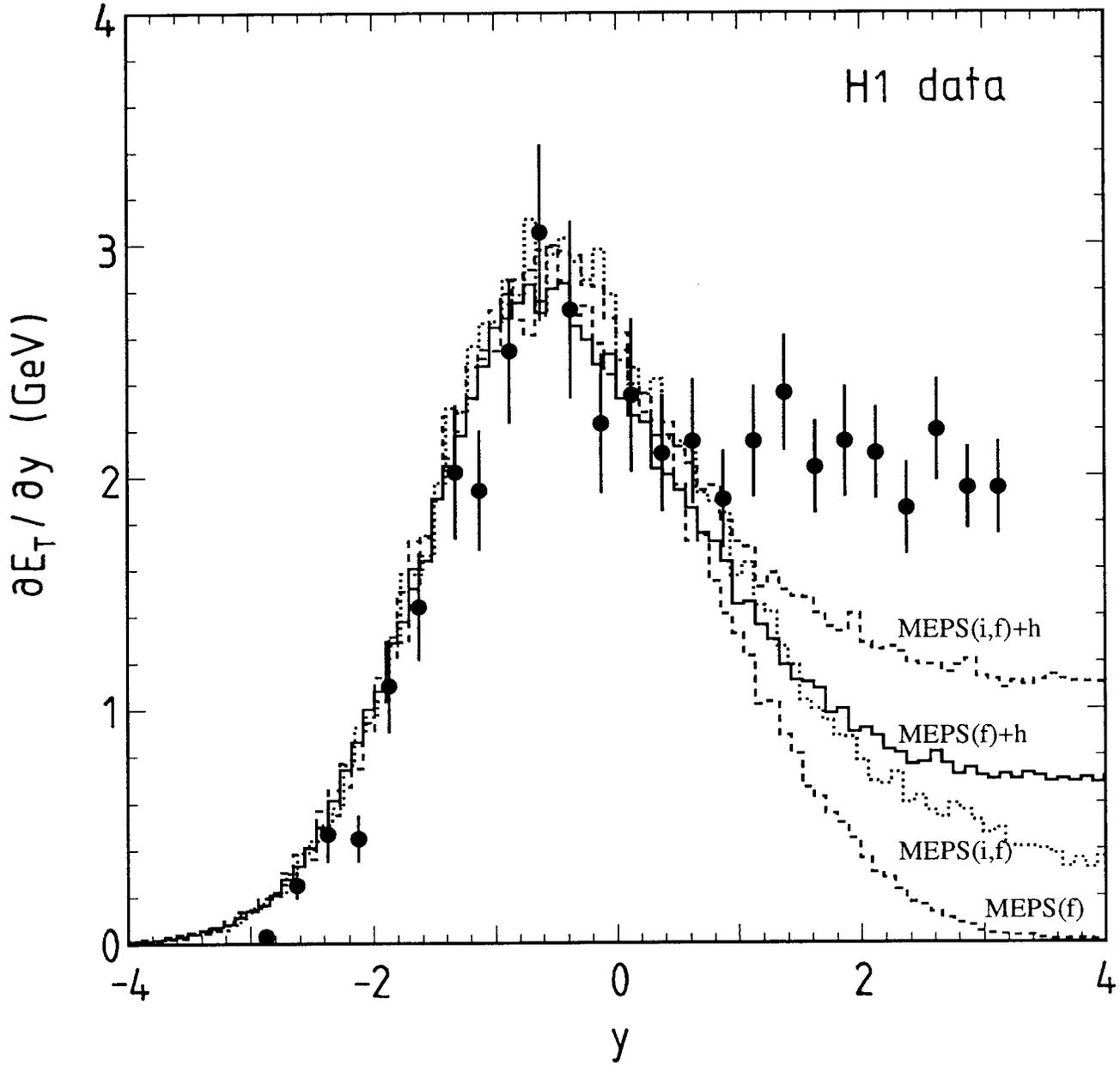
\item[Fig.\ 3:]
 The data show the $E_T$ flow as a function of rapidity 
in the laboratory (HERA)
frame which accompanies deep-inelastic events with $x<10^{-3}$
\cite{H1}.  The proton direction is to the right.  The LEPTO
Monte Carlo distributions correspond, in increasing order, to
ME+PS(final), ME+PS(initial,final),
ME+PS(final)~+ ha\-dro\-ni\-za\-tion, and finally
ME+PS(initial,final)+ ha\-dro\-ni\-za\-tion.

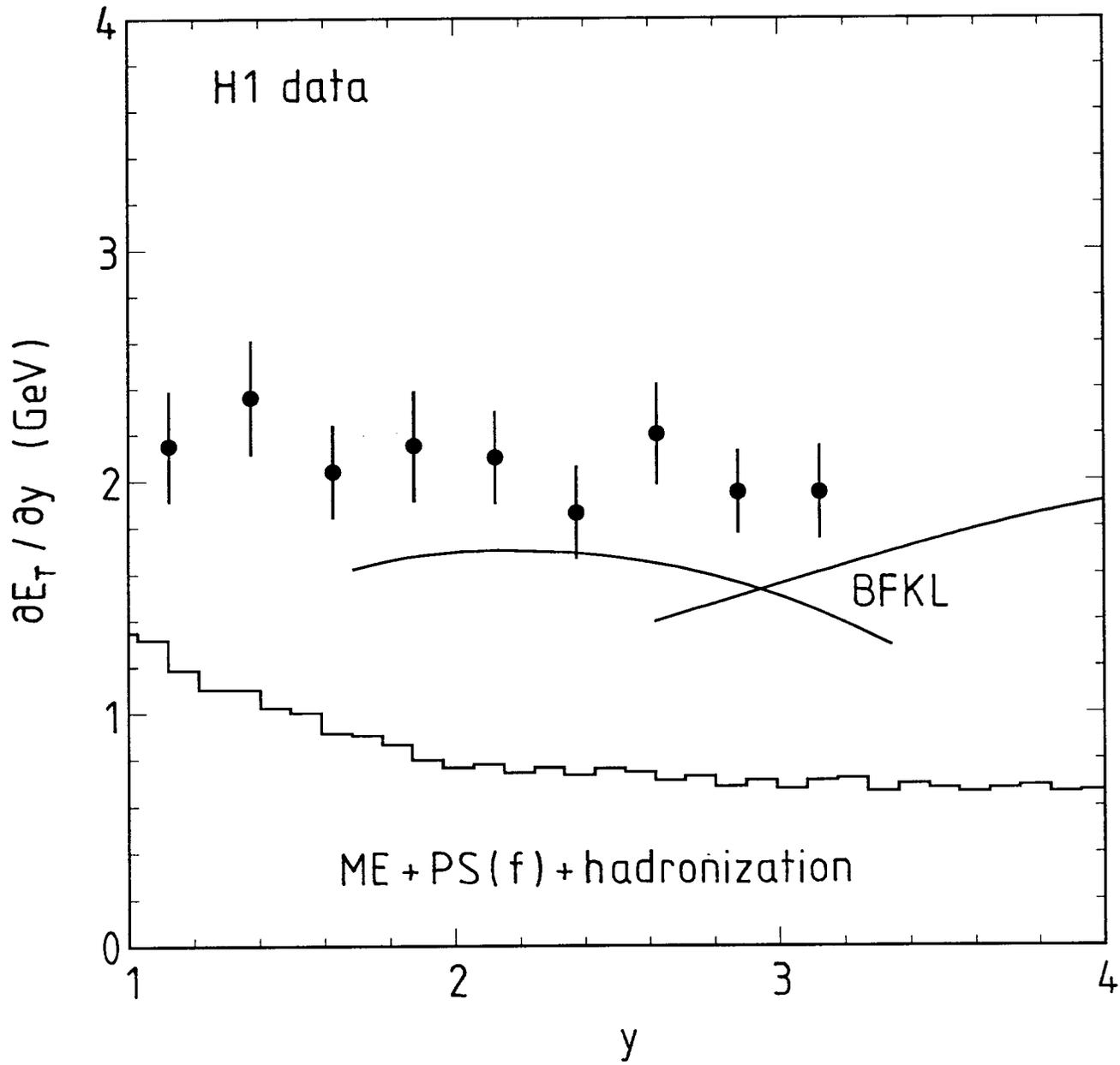
\item[Fig.~4:]
The data show the $E_T$ flow accompanying deep-inelastic events with
$x<10^{-3}$ observed by the H1 collaboration \cite{H1} in the central region.
The continuous curves show the BFKL predictions of $x=5.7 \times
10^{-4}$ and $Q^2 = 15$ GeV$^2$, which correspond to the average
values of the variables for the data sample. The histogram is the
LEPTO Monte Carlo estimate from Fig.~3 of the effects of radiation
from the current jet and of hadronization. 
\end{itemize}
\end{document}